\documentclass[12pt,a4paper,twoside]{paper}
\usepackage{amsmath,bbm,amssymb,amsxtra}
\usepackage{graphicx}
\usepackage{epstopdf}
\usepackage{chngcntr}
\usepackage{etoolbox}

\usepackage{bbold}

\usepackage{fancyhdr}
 \usepackage{fancyhdr}
\newcommand{\be}{\begin{equation}}
\newcommand{\ee}{\end{equation}}

\newcommand{\ba}{\begin{eqnarray}}
\newcommand{\ea}{\end{eqnarray}}

\newcommand{\bea}{\begin{eqnarray*}}
\newcommand{\eea}{\end{eqnarray*}}

\newcommand{\bee}{\begin{enumerate}}
\newcommand{\ene}{\end{enumerate}}

\usepackage{epsfig}
\def\R{\mathbb R}

\def\Z{\mathbb Z}

\usepackage{appendix}
\usepackage{chngcntr}
\usepackage{etoolbox}
\usepackage{lipsum}
\usepackage{amsmath,bbm,amssymb,amsxtra}
\usepackage{enumerate}
\allowdisplaybreaks
\usepackage[active]{srcltx}
\usepackage{epsfig}
\usepackage{ae}
\numberwithin{equation}{section}

\newtheorem{lemma}{Lemma}[section]
\newtheorem{proposition}[lemma]{Proposition}

\newtheorem{theorem}[lemma]{Theorem}

\newtheorem{definition}[lemma]{Definition}

\newcommand{\prop}[1]{\begin{proposition}\label{#1}
\sl }
\newcommand{\eprop}{\end{proposition}}
\newcommand{\thm}[1]{\begin{theorem}\label{#1}
\Ä }
\newcommand{\ethm}{\end{theorem}}
\newcommand{\lem}[1]{\begin{lemma}\label{#1}
\sl }
\newcommand{\elem}{\end{lemma}}

\newcommand{\defin}[1]{\begin{definition}\label{#1}
\sl }
\newcommand{\edefin}{\end{definition}}

\newcommand{\sgn}{{\rm sgn}\,}

\def\la{\lambda}

\def\CA{{\mathcal A}}       \def\CB{{\mathcal B}}       
              
       \def\CH{{\mathcal H}}

\def\CH{{\mathcal H}}

\def\qq{ \begin{eqnarray} }
\def\qqq{ \end{eqnarray} }
\def\rr{ \begin{equation} }
\def\rrr{ \end{equation} }
\def\non{ \nonumber }
\def\qq{ \begin{eqnarray} }
\def\qqq{ \end{eqnarray} }
\def\non{ \nonumber }

\newcommand{\p}{\partial}

\usepackage{epsfig}

\counterwithin*{equation}{section} 
\counterwithin*{equation}{subsection} 

\ifx\optionkeymacros\undefined\else \fi

\catcode`\Œ=\active\defŒ{{\aa}}       
\catcode`\º=\active\defº{\int}        
\catcode`\=\active\def{\c c}        
\catcode`\¶=\active\def¶{\partial}    
\catcode`\Ä=\active\defÄ{\oint}       
\catcode`\Æ=\active\defÆ{\triangle}   
\catcode`\Â=\active\defÂ{\neg}        
\catcode`\µ=\active\defµ{\mu}         
\catcode`\¿=\active\def¿{{\o}}        
\catcode`\¹=\active\def¹{\pi}         
\catcode`\Ï=\active\defÏ{{\oe}}       
\catcode`\§=\active\def§{{\ss}}       
\catcode`\ =\active\def {\dagger}     
\catcode`\Ã=\active\defÃ{\sqrt}       
\catcode`\·=\active\def·{\Sigma}      
\catcode`\Å=\active\defÅ{\approx}     
\catcode`\½=\active\def½{\Omega}      
\catcode`\£=\active\def£{{\it\$}}     
\catcode`\°=\active\def°{\infty}      
\catcode`\¤=\active\def¤{{\S}}        
\catcode`\¦=\active\def¦{{\P}}        
\catcode`\¥=\active\def¥{\bullet}     
\catcode`\»=\active\def»{\leavevmode\raise.585ex\hbox{\b a}}      
\catcode`\¼=\active\def¼{\leavevmode\raise.6ex\hbox{\b o}}        
\catcode`\­=\active\def­{\not=}       
\catcode`\²=\active\def²{\leq}        
\catcode`\³=\active\def³{\geq}        
\catcode`\Ö=\active\defÖ{\div}        
\catcode`\É=\active\defÉ{{\dots}}     
\catcode`\¾=\active\def¾{{\ae}}       
\catcode`\Ç=\active\defÇ{\ll}         
\catcode`\Ò=\active\defÒ{``}          
\catcode`\Á=\active\defÁ{!`}          
\catcode`\¢=\active\def¢{\rlap/c}     
\catcode`\Ô=\active\defÔ{`}           
\catcode`\Õ=\active\defÕ{'}           

\catcode`\=\active\def{{\AA}}       
\catcode`\'=\active\def'{\c C}        
\catcode`\¯=\active\def¯{{\O}}        
\catcode`\¸=\active\def¸{\Pi}         
\catcode`\Î=\active\defÎ{{\OE}}       
\catcode`\®=\active\def®{{\AE}}       
\catcode`\×=\active\def×{\diamond}    
\catcode`\¡=\active\def¡{\accent'27}  
\catcode`\Ó=\active\defÓ{''}          
\catcode`\±=\active\def±{\pm}         
\catcode`\È=\active\defÈ{\gg}         
\catcode`\À=\active\defÀ{?`}          
\catcode`\Ð=\active\defÐ{--}          
\catcode`\Ñ=\active\defÑ{---}         

\catcode`\Š=\active\defŠ{\"a}        
\catcode`\'=\active\def'{\"e}        
\catcode`\•=\active\def•{\"{\i}}     
\catcode`\š=\active\defš{\"o}        
\catcode`\Ÿ=\active\defŸ{\"u}        
\catcode`\Ø=\active\defØ{\"y}        
\catcode`\€=\active\def€{\"A}        
\catcode`\…=\active\def…{\"O}        
\catcode`\†=\active\def†{\"U}        
\catcode`\‡=\active\def‡{\'a}        
\catcode`\Ž=\active\defŽ{\'e}        
\catcode`\'=\active\def'{\'{\i}}     
\catcode`\—=\active\def—{\'o}        
\catcode`\œ=\active\defœ{\'u}        
\catcode`\ƒ=\active\defƒ{\'E}        
\catcode`\ˆ=\active\defˆ{\`a}        
\catcode`\=\active\def{\`e}        
\catcode`\"=\active\def"{\`{\i}}     
\catcode`\˜=\active\def˜{\`o}        
\catcode`\=\active\def{\`u}        
\catcode`\Ë=\active\defË{\`A}        
\catcode`\‹=\active\def‹{\~a}        
\catcode`\–=\active\def–{\~n}        
\catcode`\›=\active\def›{\~o}        
\catcode`\Ì=\active\defÌ{\~A}        
\catcode`\"=\active\def"{\~N}        
\catcode`\Í=\active\defÍ{\~O}        
\catcode`\‰=\active\def‰{\^a}        
\catcode`\=\active\def{\^e}        
\catcode`\"=\active\def"{\^{\i}}     
\catcode`\™=\active\def™{\^o}        
\catcode`\ž=\active\defž{\^u}        

\let\optionkeymacros\null

\begin{document}


     \begin{center}
 \textsl{\LARGE    EPR-Bell-Schrödinger proof of nonlocality using position and momentum}\footnote{To appear in: Do wave functions jump? Perspectives on the work of G.C. Ghirardi. Editors: V. Allori, A. Bassi, D. Dürr,  N. Zanghì.}

 \vspace*{5mm}
 { \Large Jean Bricmont
 
 IRMP,
Universit\'e catholique de Louvain,
chemin du Cyclotron 2,
1348 Louvain-la-Neuve
Belgium}\footnote{E-mail: jean.bricmont@uclouvain.be}
 \vspace*{5mm}
 
{ \Large Sheldon Goldstein

 Department of Mathematics, Rutgers University, Hill Center, 110 Frelinghuysen Road, Piscataway,
NJ 08854-8019, USA}\footnote{E-mail: oldstein@math.rutgers.edu}

 \vspace*{5mm}
{ \Large Douglas Hemmick

 66 Boston Drive,
Berlin, MD 21811, USA}\footnote{E-mail: jsbell.ontarget@gmail.com}

\end{center}

\vspace*{10mm}
  { \Large  We dedicate this paper to the memory of Giancarlo Ghirardi, who devoted his life to understanding 
  quantum mechanics. He  was a friend of John Bell, who was inspired by Giancarlo's work. He  was also a friend of two of us (J.B. and S.G.).}

\begin{abstract}

 Based on his extension  of the classical argument of Einstein, Podolsky and Rosen, Schrödinger observed that, in certain quantum states associated with pairs of particles that can be far away from one another, the result of the measurement of an observable associated with one particle is perfectly correlated with the result of the measurement of another observable associated with the other particle. Combining this with the assumption of locality and some ``no hidden variables" theorems, we showed in a previous paper \cite{1} that this yields a contradiction. This means that the assumption of locality is false, and thus provides us with another demonstration of quantum nonlocality that does not involve Bell's (or any other) inequalities. In \cite{1} we introduced only ``spin-like" observables acting on finite dimensional Hilbert spaces. Here we will give a similar argument using the variables originally used by Einstein, Podolsky and Rosen, namely position and momentum.

\end{abstract}

\section{Introduction}\label{sec1}

In 1935, Einstein, Podolsky and Rosen (EPR) argued that quantum mechanics is incomplete by considering two particles in one dimension moving in opposite directions and whose joint wave function (see (\ref{Psi_3}) below) was such that the measurement of the position of one of the particles immediately determined the position of the other particle  and, similarly, the measurement of the momentum of one of the particles immediately determined the momentum of the other one. 

Since, said EPR, a measurement made on one particle obviously could not possibly influence the physical state of the other particle, situated far away from the first particle, and since the wave function of both particles
specifies neither the position nor the momentum of those particles, this quantum mechanical description of the state of both particles provided by this wave function must be {\it incomplete} in the sense that other variables, such as the values of the positions and momenta of both particles, must be included in a complete description of that physical system.

EPR's argument had been widely misunderstood and misrepresented or ignored by almost everybody at that time. But not by Schr\"odinger, who, in his ``cat paper," originally published in German \cite{Sch}, as well as in the papers \cite{Sch1, Sch2}, understood the ``paradox" raised by EPR and deepened the perplexity  that it causes.

Schr\"odinger showed that for certain states, called now maximally entangled (see Subsection \ref{sec2.1}),
it is not just that the positions and the momenta of the particles are perfectly correlated. He showed that, for every observable associated with the first particle, there is another observable associated with the second particle such that the results of the measurements of both observables are perfectly correlated.

In \cite{1}, following \cite{Hem, Hem-S}, we explained that, if one assumes locality, meaning that there is no effect whatsoever on the state of the second particle due to a measurement carried out on the first particle (when both particles are sufficiently spatially separated), there must exist what we call a ``non-contextual value map" $v$ which assigns to each observable $A$ a value $v(A)$ that pre-exists its measurement and is simply revealed by it. The word ``non-contextual" refers to the fact that, since it pre-exists the measurement, the value $v(A)$ does not depend on the procedure used to measure $A$.

However several theorems, originally due to Bell \cite{Be1} and to Kochen and Specker \cite{KS}, preclude the possibility of a  non-contextual value map.\footnote{In the literature on quantum mechanics, these theorems  are often called ``no hidden variables" theorems. But we prefer the expression ``inexistence of a non-contextual value map" because, as we will discuss in Section \ref{sec5}, the expression ``hidden variables" is really a misnomer.} Since the existence of this map is a logical consequence of  the assumption of locality and of the  perfect correlations, the assumption of locality is false.

In this paper, we first summarize the arguments of our previous paper \cite{1} (Section \ref{sec2}). We then turn to the EPR paper as well as related work by  Einstein alone and  Schr\"odinger  (Section \ref{sec3}). Next, we  provide a  proof of nonlocality similar to the one  of Section \ref{sec2}, but
using only functions of the EPR variables, namely positions and momenta (Section \ref{sec4}). This argument relies on a theorem of Robert Clifton \cite{RC}. 

We then consider what happens in  Bohmian mechanics (Section \ref{sec5}): in that theory,  particles have, at all times, both a position and a momentum and  one might therefore think that this would imply the existence of a   non-contextual value-map  for functions of those variables. We explain however, through an analysis of what a measurement of momentum means in that theory, that this is not the case. Finally we briefly discuss how nonlocality manifests itself in Bohmian mechanics.

 For a discussion of the relationship between this work and previous 
ones, including \cite{H-R, St, Brown, El, Cab, Ara}, see  Section 7 of 
\cite{1}.

\section{Proof of nonlocality based on perfect correlations}\label{sec2}

We will first discuss special quantum states, called maximally entangled, for pairs of physical systems that can possibly be located far apart, and having the property that, for each quantum observable of one of the systems, there is an associated observable of the other one such that the result of the measurement of that observable is perfectly correlated with the result of the measurement on the first one.
  
\subsection{Maximally entangled states}\label{sec2.1}

Consider a finite dimensional (complex) Hilbert space $\cal H$, of dimension $N$,   and orthonormal bases $\psi_n$ and $\phi_n$ in $\cal H$ (we will assume below that all bases are orthonormal).
A unit vector $\Psi$ in $\cal H \otimes \cal H$ is {\it maximally entangled} if it is of the form 
\be
\Psi= \frac{1}{\sqrt N}\sum_{n=1}^N    \psi_n \otimes \phi_n.
\label{ME}
\ee

 Since we are interested in quantum mechanics, we will refer to those vectors as {\it maximally entangled states} and we will associate, by convention, each space in the tensor product with a ``physical system," namely we will consider the set  $\{\phi_n\}_{n=1}^N$ as a basis of states for physical system 1  and the set  $\{\psi_n\}_{n=1}^N$ as a basis of states for physical system 2.

Now, given a  maximally entangled state, one can  associate to each operator of the form ${\mathbb 1}  \otimes O$ (meaning that it acts non-trivially only on particle 1) an operator of the form $ \tilde O \otimes  {\mathbb 1}  $ (meaning that it acts non-trivially only on particle 2). Here ${\mathbb 1}$ denotes the identity operator on  $\cal H$.

Define the operator $U$ mapping $\cal H$ to $\cal H$ by setting
\be
U \phi_n =\psi_n,
\label{defU}
\ee
$\forall n= 1, \dots, N$, and extending $U$ to an anti-linear operator on all of $\cal H$:
\be
U (\sum_{n=1}^N c_n \phi_n) =\sum_{n=1}^N c^*_n U \phi_n= \sum_{n=1}^N c^*_n \psi_n
\label{U1}
\ee
where $^*$ denotes the complex conjugate.

Using the operator $U$, the state $\Psi$ in (\ref{ME}) can be written as:
\be
\Psi= \frac{1}{\sqrt N}\sum_{n=1}^N   U\phi_n \otimes \phi_n .
\label{ME1}
\ee
It is easy to check that this formula is the same for any basis, see \cite[eq. 3.1.8]{1}. 
 
$U$ thus determines, and is uniquely determined by, a maximally entangled state $\Psi$.

Given such a state $\Psi$, and hence $U$, we may associate
 to every operator of the form  ${\mathbb 1}  \otimes O$ an operator of the form  $ \tilde O \otimes {\mathbb 1}$ by setting
\be
\tilde O = U O  U^{-1}.
\label{A}
\ee

Suppose $\phi_n$ are eigenstates of $O$, with eigenvalues $\la_n$,
\be
 O \phi_n = \la_n \phi_n.
\label{A1}
\ee
Then, the states $\psi_n= U \phi_n$ are eigenstates of $\tilde O$, also with eigenvalues $\la_n$:
\be
\tilde O \psi_n = \la_n \psi_n.
\label{A2}
\ee
This implies and is in fact equivalent to the following relationship between the operators $O$ and $\tilde O$:
\be
( O \otimes  {\mathbb 1} - {\mathbb 1} \otimes \tilde O ) \Psi = 0,
\label{A3}
\ee
directly expressing the fact that, in the state $ \Psi$, $O \otimes  {\mathbb 1}$ and ${\mathbb 1} \otimes \tilde O$
are perfectly correlated.

We  may summarize this as follows:
\begin{theorem}\label{0}
Consider a finite dimensional  Hilbert space $\cal H$, of dimension $N$,  and a  maximally entangled state $\Psi \in \cal H \otimes \cal H$. Then, for any
self-adjoint operator $O$ acting on $\cal H$,
 there exists a self-adjoint operator  $\tilde O$ acting on $\cal H$ such that (\ref{A3}) holds.
\end{theorem}

{\bf Remarks}

\begin{itemize}

\item[1.]
A simple example of a maximally entangled state is:
\begin{eqnarray}
|\Psi\rangle &=& \frac{1}{\sqrt 2} \big(|  \uparrow  \rangle| \downarrow  \rangle-|  \downarrow  \rangle|   \uparrow  \rangle\big),
\label{1}
\end{eqnarray}
where the right factors refer to system $1$ and  left ones  to system $2$.
That state, according to ordinary quantum mechanics, means that the spin measured on  system $1$ will have equal probability to be up or down, but is perfectly anti-correlated with the spin measured on system $2$.

 In the notation of (\ref{ME}), one has:
 \be
\nonumber
\phi_1= | \uparrow>,\;\; \phi_2 = | \downarrow>, \;\;\psi_1= -|\downarrow>, \;\; \psi_2=|\uparrow>,
\ee
and therefore,
\ba
\nonumber U |  \uparrow>&=& -|  \downarrow>,\\\nonumber
U |  \downarrow>&=& |  \uparrow>.
\ea
If one takes 
\ba
O = \left(\begin{array}{ccc} 1 & 0 \\ 0 & -1 \end{array}\right)
\ea
which corresponds to the spin operator for system $1$ and  has eigenvectors $\phi_1$ with eigenvalue $1$ and $\phi_2$ with eigenvalue $-1$, one computes that
\ba
\tilde O =  U O  U^{-1}= \left(\begin{array}{ccc} -1 & 0 \\ 0 & 1 \end{array}\right)=-O,
\ea
which means that the spin operator for systems $1$ and $2$ are perfectly anti-correlated, since $\tilde O$ is minus the spin operator for system $2$.

We will use later the following:  

\item[2.] 

{\it Products of maximally entangled states are maximally entangled states}: If one has two Hilbert spaces $ {\cal H}_1$, $ {\cal H}_2$, and two   maximally entangled states $\Psi_i \in {\cal H}_i \otimes {\cal H}_i$, $i= 1, 2$, then it is easy to check that the state $\Psi= \Psi_1 \otimes \Psi_2$ is maximally entangled  in ${\cal H} \otimes {\cal H}$, where
${\cal H}= {\cal H}_1 \otimes {\cal H}_2$ (under the canonical identification of $({\cal H}_1 \otimes {\cal H}_1) 
\otimes ({\cal H}_2 \otimes {\cal H}_2)$ with $ {\cal H}\otimes {\cal H}$).

\end{itemize}

Let us now see what this notion of maximally entangled state implies for quantum measurements.

Suppose that we have a pair of physical systems, whose states belong to the same finite dimensional  Hilbert space $\cal H$. And suppose that the quantum state $\Psi$ of the pair is maximally entangled, i.e. of the form (\ref{ME}).

Any observable acting on system 1 is represented by a self-adjoint operator $O$, which has therefore a basis of eigenvectors. Since the representation (\ref{ME1}) of the state $\Psi$ 
is valid in any basis, we may choose, without loss of generality, as the set $\{\phi_n\}_{n=1}^N$  in (\ref{ME})  the eigenstates of $O$. Let  $\la_n$  be the corresponding  eigenvalues, see  (\ref{A1}). 

 If one measures that observable $O$, the result will be one of the eigenvalues $\la_n$, each having equal probability $\frac{1}{N}$. If the result is $\la_k$,  the (collapsed) state of the system after the measurement, will be $\psi_k \otimes \phi_k $. Then, the measurement of observable $\tilde O$, defined by (\ref{A}, \ref{defU}), on system 2, will necessarily yield the value $\la_k$. 
 
Reciprocally, if one measures an observable $\tilde O$ on system 2 and the result is  $\la_l$,  the (collapsed) state of the system after the measurement, will be $\psi_l \otimes \psi_l $, and the measurement of observable $ O$ on system 1 will necessarily yield the value $\la_l$. 

To summarize, we have derived the following consequence of the quantum formalism:

{\bf Principle of Perfect Correlations.}
{\it In any maximally entangled  quantum state, of the form (\ref{ME}), there is, for each operator $O$ acting on system 1, an operator $ \tilde O$ acting on system 2 (defined by (\ref{A}, \ref{defU})), such that, if one measures the physical quantity represented by operator $ \tilde O$ on system 2 and the result is the eigenvalue  $\la_l$ of $ \tilde O$, then, measuring the physical quantity represented by operator 
$O$  on system 1 will yield with certainty the same eigenvalue $\la_l$, and vice-versa.\footnote{The correlations mentioned here are often called  anti-correlations, for example when $\tilde O=-O$, as in the example of the spin in remark 1 above.}}

\subsection{Schr\"odinger's ``Theorem"}\label{sec2.2}

The following property  will be  crucial in the rest of the paper.
 
{\bf Locality.}
{\it If systems 1 and 2 are spatially separated from each other, then measuring an observable on system 1 has no instantaneous effect whatsoever on system 2 and  measuring an observable on system 2 has no instantaneous effect whatsoever on system 1.}

Finally, we must also define: 
 
 {\bf Non-contextual value-maps.}
 Let $\cal H$ be  a finite dimensional  Hilbert space  and let $\CA$ be the set of self-adjoint operators on $\CH$. Suppose $\cal H$ is the quantum state space for a physical system and $\CA$ is the set of quantum observables. Suppose there are situations in which there are observables $A$ for which the result of measuring $A$ is determined already, before the measurement. Suppose, that is, that $A$ has, in these situations, a pre-existing value $v(A)$ revealed by measurement and not merely created by measurement. Of course, this implies that for every experiment ${\cal E}_A$ measuring $A$, the result $v({\cal E}_A)$ of that experiment, in the situation under consideration, must be $v(A)$. And suppose finally that the situation is such that we have a pre-exiting value $v(A)$ for every $A\in \CA$. 
 
 We would then have a 
 {\it  non-contextual value-map}, namely  a map  $v: \CA \to \R$ that assigns the value $v(A)$ to  any experiment associated with what is called in quantum mechanics a measurement of an observable $A$. There can be different ways to measure the same observable. The  value-map is called non-contextual  because all such experiments, associated with the same quantum observable $A$, are assigned the same value.

This notion of value-map is not a purely mathematical one, since it involves the notion of an experiment that measures a quantum observable $A$, which we have not mathematically formalized. However, we shall need only the following obvious purely mathematical consequence of non-contextualitty.

  A  non-contextual value-map has the fundamental property that, if $A_i$, $i=1, \dots, n$, are mutually commuting self-adjoint operators on $\CH$, $
 [A_i, A_j]= 0, \forall i, j =1, \dots, n$, then, if $f$ is a  function of $n$ variables and $B= f(A_1, \dots, A_n)$, then
 
 \be
v(B)= f(v(A_1), \dots, v(A_n)).
\label{res}
\ee

It is a well-known property of quantum mechanics that, since all the operators $A_1, \dots, A_n, B$ commute, they are simultaneously measurable and the result of those measurements must satisfy (\ref{res}).

But, and this is what we  emphasized in \cite{1}, (\ref{res}) follows trivially from the non-contextualilty of the value-map. Indeed, a valid quantum mechanical  way to measure the operator $B= f(A_1, \dots, A_n)$ is to measure $A_1, \dots, A_n$ and, denoting the results $\la_1, \dots, \la_n$, to regard $\la_B=f(\la_1, \dots, \la_n)$ as the result of a measurement of $B$ .
Since, by the non-contextuality of the map $v$, all the possible measurements of $B$ must yield the same results, (\ref{res}) holds.

Thus, once one has a non-contextual value-map,
 {\it one does not even need to check} (\ref{res}).

Now we will use the perfect correlations and locality to establish the existence of a non-contextual value-map $v$,
for a maximally entangled  quantum state of the form (\ref{ME}) or, equivalently, (\ref{ME1}). By the 
principle of perfect correlations,
or any  operator $O$ on system 1, there is an  operator  $\tilde O$  on system 2, defined by (\ref{A}, \ref{defU}), which  is perfectly correlated with $O$ through (\ref{A3}).

Thus, if we were to measure  $\tilde O$, obtaining  $\la_l$, we would know that 
\ba
v(O)= \la_l 
\label{map}
\ea
concerning the result of then measuring $O$. Therefore, $v(O)$ would pre-exist the measurement of $O$.
But, by the assumption of locality, the measurement of   $\tilde O$, associated with the second system,  could not have 
had any effect on the first system,
and thus, this value $v(O)$ would pre-exist also  the measurement of  $\tilde O$ and this would not depend upon whether $\tilde O$ 
had been measured.
 Letting $O$ range over all operators on system 1,  
we see that there must be a non-contextual value-map $O\to v(O)$. 

To summarize, we have shown:

 \noindent
{\bf Schr\"odinger's ``Theorem".} Let $\CA$ be the set of self-adjoint operators on the component Hilbert space   $\CH$ of a physical system 
in a maximally entangled state (\ref{ME}). Then, assuming locality and the principle of perfect correlations, there 
exists a non-contextual value-map  $v: \CA \to \R$.

{\bf Remark}

\begin{itemize}
\item[] 
We put ``Theorem" in quotation marks because the  statement concerns physics and not just mathematics. Its conclusions are nevertheless inescapable  assuming the hypothesis of locality and the empirical validity of the principle of perfect correlations, a principle which is, as we showed, a consequence of the quantum formalism.

 \end{itemize}

\subsection{The non-existence of non-contextual value-maps}\label{sec2.3}

The problem posed by the non-contextual value-map $v$ whose existence is implied by Schr\"odinger's ``theorem" is that such maps simply do not exist (and that is a purely mathematical result). Indeed, one has the:

\noindent
{\bf ``Theorem": Non-existence of non-contextual value-maps.}
Let $\CA$ be the set of self-adjoint operators on the  Hilbert space $\cal H$ of a physical system. Then there exists no non-contextual value-map $v: \CA \to \R$.
 
This ``theorem" is an immediate consequence of the following theorem, since (\ref{res2}, \ref{res3}) are consequences of (\ref{res}).\footnote{This is obvious for (\ref{res3}),  a special case of (\ref{res}).
 For (\ref{res2}) we observe that, since $O$
is self-adjoint, we can write 
$O = \sum_i \lambda_i P_{\lambda_i}$ where $P_{\lambda_i}$ is the projector on the subspace
of eigenvectors   of eigenvalue $\lambda_i$ of $O$
and thus we have that $f(O) = \sum_i f(\lambda_i) P_{\lambda_i}$. If we choose any $f$ whose range is the set of  eigenvalues of $O$ and is such that $f(\lambda_i)= \lambda_i$ $\forall i$, we have that 
$O=f(O)$ and, 
by (\ref{res}), we obtain that $v(O)=v(f(O))=f(v(O))$ and thus  $v(O)$ is an eigenvalue of $O$.}

\begin{theorem}\label{2}
 Let $\cal H$ be  a finite dimensional  Hilbert space of dimension at least three,  and let $\CA$ be the set of self-adjoint operators on $\CH$.
There does not  exist a map  $v: \CA \to \R$ such that:

1) $ \; \forall O \in { \CA}$, 
\be
v(O) \;\;\mbox{is an eigenvalue of} \;\;O
\label{res2}
\ee

2) $\forall O, O' \in { \CA}$ with  $ [O, O']= OO'-O'O=0$, and
for any real valued function $f$ of two real variables,
\be
v(f(O, O'))=f(v(O), v(O')).\label{res3}\\
\ee
\end{theorem}

See \cite{1} for a discussion of the proof of the theorem, which is a consequence of stronger theorems, originally due to John Bell \cite{Be1} and to Kochen and Specker \cite{KS},  with simplified proofs of Theorem \ref{2} due to David Mermin \cite{Me4}, and to Asher Peres \cite{Per, Per1}.

\subsection{Nonlocality}\label{sec2.4}

The conclusion of Schr\"odinger's ``theorem" and of the ``Theorem"  on the non-existence of non-contextual value-maps plainly contradict each other.
So, the assumptions of at least one of them must be false.  Moreover, the stronger Theorem \ref{2} is a purely mathematical result. 
To derive Schr\"odinger's ``theorem," we assume only the perfect correlations and locality. The  perfect correlations are an immediate consequence of quantum mechanics.   The only remaining assumption  is locality. Hence we can deduce:

\noindent
{\bf Nonlocality ``Theorem"}. The locality assumption is false.

See \cite[sections 5, 7]{1} for a discussion of the relation between this proof  and other proofs  of nonlocality.

\section{The Original EPR Argument}\label{sec3}

Let us now turn to the original EPR argument \cite{EPR} and explain its connection to the notion of
locality.
EPR gave both a general argument and a specific example. 

\subsection{EPR's general setup}\label{sec3.1}

For their general argument, EPR considered a system of two particles, 1 and 2, in one dimension, that may be far apart and   a physical quantity  represented by a self-adjoint operator $O$ that acts on system 1. We shall assume that $O$ has an orthonormal basis of eigenvectors $ \phi_n (x_1)$
with eigenvalues $\la_n$.

One can then write the joint state of both particles as:

\begin{eqnarray}
\Psi(x_1, x_2) &=& \sum_{n=1}^\infty \psi_n(x_2) \phi_n (x_1),
\label{Psi_1}
\end{eqnarray}
where $\psi_n(x_2)$ are the ($x_2$ dependent) coefficients of that expansion.\footnote{This ressembles a maximally entangled state, like (\ref{ME}), but it is not one because the sum in (\ref{Psi_1}) extends to infinity and, for (\ref{Psi_1}) to be a maximally entangled state, the set $\{\psi_n\}_{n=1}^\infty$ should be orthonormal. But then the norm of (\ref{Psi_1}) would be infinite and thus (\ref{Psi_1}) would not belong the the Hilbert space. 
}

After a measurement of $O$ on system 1, if the result is $\la_l$, then the state collapses to  $\psi_l(x_2) \phi_l (x_1) $, i.e. $\phi_l (x_1)$ for the first particle and $\psi_l(x_2)$
for the second.

If, on the other hand, one considers  a physical quantity  represented by an operator $O'$ that acts on system 1, and  one assumes that $O'$  has eigenvectors $ \phi'_s (x_1)$
and eigenvalues $\mu_s$, one can write the joint state as:

\begin{eqnarray}
\Psi(x_1, x_2) &=& \sum_{s=1}^\infty \psi'_s(x_2) \phi'_s (x_1)
\label{Psi_2}
\end{eqnarray}

After a measurement of $O'$ on system 1, if the result is $\mu_k$, then the state collapses to  $\psi'_k(x_2) \phi'_k (x_1)$, i.e.  $\phi'_k (x_1)$ for the first particle and $\psi'_k(x_2)$
for the second.

We will discuss the implications of that observation after giving the concrete examples of the operators considered by EPR.

\subsection{The example of position and momentum}\label{sec3.2}

For their specific example, EPR  introduced  a two particle wave function\footnote{This is a generalized wave function, which means that it is not an element of the Hilbert space $L^2(\R^2)$, but rather a distribution, namely a linear function acting on a space of smooth functions that decay rapidly  at infinity (see \cite[Section 5.3]{RS1} for a short introduction to distributions). We  will not try to be rigorous about these generalized functions here, but we will
give a regularized version of the same wave function in Subsection \ref{sec3.6}.}:

\begin{eqnarray}
\Psi_{EPR} (x_1, x_2)  &=& \int_{-\infty}^{\infty} \exp(i(x_1-x_2+x_0)p) dp
\label{Psi_3}
\end{eqnarray}
(putting $\hbar=1$).
 This can be written, by analogy with (\ref{Psi_1}), i.e. with sums replaced by integrals, as:

\begin{eqnarray}
\Psi_{EPR} (x_1, x_2)  &=& \int_{-\infty}^{\infty} \psi_p(x_2) \phi_p (x_1) dp
\label{Psi_3b}
\end{eqnarray}

with:
 $ \phi_p (x_1)= \exp(ix_1p) $, 
and
$\psi_p(x_2)= \exp(-i(x_2-x_0)p)$.

It will be useful to introduce  the Fourier transform of a wave function $\Psi$:
\ba
\widehat \Psi( p_1, p_2)= \frac{1}{2\pi} \int \exp(-i (p_1 x_1+p_2 x_2))\Psi( x_1, x_2) dx_1 dx_2,
\label{FT}
\ea
whose inverse is:
\ba
 \Psi( x_1, x_2) =  \frac{1}{2\pi} \int \exp(i (p_1 x_1+p_2 x_2))  \widehat \Psi( p_1, p_2) dp_1 dp_2.
\label{IFT}
\ea
EPR took the operator $O$ to be the momentum operator 
$$P_1=-i\frac{d}{dx_1}
$$ 
acting on the first particle and on a suitable set of functions (see \cite[Chapter VIII]{RS1} for precise definitions).

We know that $ \phi_p (x_1)= \exp(ix_1p) $ is a (generalized) eigenstate of $P_1$ of eigenvalue $p$, and $\psi_p(x_2)= \exp(-i(x_2-x_0)p)$  is a (generalized)  eigenstate of eigenvalue $-p$
of the momentum operator
$$P_2=-i\frac{d}{dx_2}
$$ 
acting on  the second particle.

Alternatively, $P_j$, $ j=1,2$, can be defined by its action on $\widehat \Psi( p_1, p_2)$:
 \ba
P_j \Psi( x_1, x_2) =  \frac{1}{2\pi} \int \exp(i (p_1 x_1+p_2 x_2))  p_j \widehat \Psi( p_1, p_2) dp_1 dp_2 \;,\quad j=1,2\;.
\label{PFT}
\ea

EPR took the operator $O'$ to be the position  operator $Q_1=x_1$ acting on the first particle.

Using a standard identity for distributions ($  \int_{-\infty}^{\infty} \exp(ixp) dp = 2\pi \delta (x)$) one can write the state (\ref{Psi_3}), as:
 
\begin{eqnarray}
\Psi_{EPR} (x_1, x_2)  
&=& 2\pi \delta(x_1-x_2+x_0)  \nonumber\\
&=& 2\pi \int_{-\infty}^{\infty} \delta(x-x_2+x_0) \delta(x_1-x) dx \nonumber\\
&=& \int_{-\infty}^{\infty} \psi'_x(x_2) \phi'_x (x_1) dx,
\label{Psi_4}
\end{eqnarray}
with $\psi'_x(x_2)= {\sqrt {2\pi}} \delta(x-x_2+x_0) $ and $\phi'_x(x_1)= {\sqrt {2\pi}}   \delta(x_1-x)$. The last formula is analogous to  (\ref{Psi_2}).

The (generalized) eigenfunctions of the operator $Q_1=x_1$ are  $\phi'_x(x_1)= {\sqrt {2\pi}}\delta(x_1-x)$, with eigenvalue $x$,
and $\psi'_x(x_2)= {\sqrt {2\pi}} \delta(x-x_2+x_0) $
is a (generalized) eigenvector of the operator $Q_2=x_2$, with eigenvalue $x+x_0$.

Therefore, depending on whether we choose to measure the operator $O$ or $O'$ on the first particle, one can produce  
two different states, $\psi_p(x_2)=  \exp(-i(x_2-x_0)p)$ and $\psi'_x(x_2)=  {\sqrt {2\pi}} \delta(x-x_2+x_0)$, for the second particle, which can be, in principle,  as far as one wants from the first one. 

Moreover, the states $\psi_p(x_2)=  \exp(-i(x_2-x_0)p)$ and $\psi'_x(x_2)= {\sqrt {2\pi}}\delta(x-x_2+x_0)$ are
  (generalized) eigenfunctions of two non-commuting operators, $P_2$ and $Q_2$.

\subsection{The conclusions of the EPR Paper by EPR}\label{sec3.3}

Since EPR assumed no actions at a distance, they concluded that the values of two non commuting observables, like $P_2$ and $Q_2$, for
the second particle, far away from where the measurements on the first particle take
place, must  have ``simultaneous reality" when the system is in the quantum state (\ref{Psi_3}). Thus, say EPR, quantum
mechanics, i.e., the description provided by the state (\ref{Psi_3}), is an incomplete description of physical reality.

But  they could have made a simpler argument: {\it considering only one variable is enough to show that quantum mechanics is incomplete}. 
  Indeed, I can {\it know }
  the position of the second particle by measuring the position of the first one. If that measurement, being made far away from the second particle, does not affect the state of the second particle, then the position of that second particle (which is left undetermined by the state (\ref{Psi_3})) must exist independently of any measurement on the first particle.

 And, since one can reason by exchanging the two particles,  one can also know the position of the first particle by measuring the one of the second particle, so that the position of the first particle must also exist independently of any measurements.
 
 Of course, they could have made the same argument about the momentum of either particle, but there was no need to bring in both quantities.

\subsection{The Conclusions of the EPR Paper By Einstein}\label{sec3.4} 
 
 In a June 19, 1935 letter to Schr\"odinger,  Einstein complained that the EPR paper had been written by Podolsky ``for reasons of language" and that the main point ``was buried, so to speak, by erudition" (\cite{Lett2}).
 
 Then Einstein explains what is, for him, the main point: in the notation used here, see (\ref{Psi_1}), if one measures 
quantity  $O$ on system 1,  the state collapses to some state  $\psi_l(x_2)$
for the second particle.
Similarly, if one measures a quantity  $O'$ on system 1,  see (\ref{Psi_2}), the state collapses  to some {\it different} state  $\psi'_k(x_2)$
for the second particle.

For the state $\Psi_{EPR}$, (\ref{Psi_3b}, \ref{Psi_4}) one obtains either a state of the form   $\psi_p(x_2)= \exp(-i(x_2-x_0)p)$, if one measures the momentum of the first particle or  a state of the form   $\psi'_x(x_2)=  {\sqrt {2\pi}}\delta(x-x_2+x_0)$, if one measures the position of the first particle.

 The fact that one  can obtain {\it two different states}  for the second particle by acting on the first particle, far away from the second one, proves that the wave function description in quantum mechanics is incomplete (assuming of course locality) since a more complete description would be provided by both states together. 

Einstein said that ``he could not care less" \cite[p. 38]{Fi} about the fact that those states, $\psi_p(x_2)= \exp(-i(x_2-x_0)p)$ and $\psi'_x(x_2)=  {\sqrt {2\pi}}\delta(x-x_2+x_0)$,
  are or are not eigenstates of some observable (related to the second particle).  

This is indeed  different, and simpler, than the conclusion of the EPR paper, but it is still more complicated than the argument that we gave in Subsection \ref{sec3.3}.

\subsection{Schrödinger's extension of EPR }\label{sec3.5} 

What Schrödinger did in his 1935 paper\footnote{This paper remained famous for his example of the cat that is ``both dead and alive", but that example will not concerned us here.} \cite{Sch} , and in \cite{Sch1, Sch2},  was to reflect on the EPR  paper \cite{EPR}. He introduced what we call here maximally entangled states and concluded that the value of every observable $O$ for the first system can be determined by  the measurement of the corresponding observable $\tilde O$ on the  second system, distant from the first one. That puzzled him a lot. Of course, like EPR, Schrödinger always assumed locality.

To illustrate his puzzlement, Schrödinger used the following example. Let $O$ be the energy of the harmonic oscillator, $O= \frac{1}{2} (
p^2+\omega^2 x^2)$ with $p=-i \frac{d}{dx}$. It is well known that the eigenvalues of the operator $O$ are of the form $\omega(n+\frac{1}{2})$, $n= 0, 1, 2, \dots$. But, argued Schrödinger,  if those values can be determined by measuring a similar operator $\tilde O$ acting on a distant system, they must pre-exist the measurement of $ O$, and that should hold true {\it for every value of $\omega$}. But, by the EPR reasoning, the values of the position $x$ and the momentum $p$ of the first system can also be determined by measuring either the operator $ \tilde x$ or the operator $\tilde p$ on the second system, so the values of $x$ and $p$ must also pre-exist  their measurements. But it is impossible for the quantity  $\frac{1}{2} (p^2+\omega^2 x^2)$ to belong to the set $\{ \omega(n+\frac{1}{2}) | n= 0, 1, 2, \dots\}$, for arbitrary values of $\omega$ and any given values of
$x$ and $p$. 

It is interesting to compare Schrödinger's attitude to that of von Neumann a little before 1935 \cite{VN} (von Neumann's book was published in German in 1932 but translated into English only in 1955); von Neumann proved a ``no hidden variable theorem" similar in its conclusion to our theorem \ref{2}, but by making the much stronger assumption that (\ref{res3}) holds even for non-commuting operators $O$
and $O'$, at least for the function $f(x,y)= x+y$, and he concluded that the  ``value-map" cannot exist. If one assumes that (\ref{res3}) holds for non-commuting operators, then it is very simple to prove the non-existence of a value-map. Take $O=\frac{1}{\sqrt 2} \sigma_x$, $O'= \frac{1}{\sqrt 2} \sigma_y$, 
where $\sigma_x$ and $\sigma_y$ are the  Pauli matrices  corresponding to the spin along the  $x$ and $y$ axes. Then $O+O'= \frac{\sigma_x + \sigma_y}{\sqrt 2} $ corresponds to the spin at an angle of $45^\circ$ between the  $x$ and $y$ axes. All the Pauli matrices  have eigenvalues equal to $\pm 1$ and so does $O+O'$. Thus $v(O)= v(O') = \pm  \frac{1}{\sqrt 2}$, and we have  $v(O)+ v(O') = \pm {\sqrt 2}$ or $0$.
But  we also have  $v(O+O')= v\big((\sigma_x + \sigma_y)/{\sqrt 2}\big)= \pm 1$. Thus equation (\ref{res3}) cannot hold for this choice of $O$ and $O'$ and $f(x,y)= x+y$. 

If Schrödinger had reasoned like von Neumann he would also have derived a ``no hidden variable theorem," using his example of the harmonic oscillator: Indeed, if $O = \frac{1}{2} (p^2+\omega^2 x^2)$, and one applies (\ref{res3}) even to non-commuting operators, one gets 
$v(O)= \frac{1}{2} (v(p)^2+\omega^2 v(x)^2)= \omega(n+\frac{1}{2})$ for some $n= 0, 1, 2, \dots$, which, as  Schrödinger observed, would be impossible for arbitrary $v(p)$, $v(x)$  and $\omega$. But Schrödinger's goal was {\it not} to prove that a value-map was impossible, since the point of  his ``theorem" was to show  that it existed (assuming locality of course). He was just baffled by the situation: recognizing that this relationship between values suggested by the form of $O = \frac{1}{2} (p^2+\omega^2 x^2)$ could not always hold, he wondered what relationship, if any, might exist among  the relevant values.
Of course, had Schrödinger made the (unwarranted) assumption of von Neumann and applied (\ref{res3}) to non-commuting operators, he would have been even more baffled, since he would probably have been led to question the locality assumption.

 Finally, note that  in 1966, much later than 1935,  John Bell constructed in \cite{Be1}  an explicit counter-example to von Neumann's conclusions, by giving a simple example of a ``hidden variables theory" that reproduces the quantum mechanical results {\it for a single spin operator} (but,  of course, without satisfying
 (\ref{res3})  for non-commuting operators). Bohmian mechanics (see Section \ref{sec5}) also provides a counter-example to von Neumann's conclusions, but a more comprehensive one.

\subsection{A regularized EPR state}\label{sec3.6}
 
A way to avoid dealing with generalized functions or distributions such as  (\ref{Psi_3}, \ref{Psi_4}) is to put a cutoff both in the spatial and the momentum variables, $x$ and $p$. A convenient way to do that is to require that $x$ take values in a finite (but arbitrarily large) 
 box on a lattice of (arbitrarily small) spacing $a$, which amounts to putting a cutoff in the momentum variable $p$.
 
 So, let $x\in \Lambda_a=[-L, L] \cap a\Z$, or $x= na, n \in \Z,  |n| \leq M$,with $M= [\frac{L}{a}]$, and  $[\cdot]$ denoting the integer part.
 
 Let 
 $\widehat \Lambda_a$ be the dual of $\Lambda_a$: 
 $$\widehat \Lambda_a= \{p= \frac{2\pi k}{a(2M+1)}, k \in \Z, |k| \leq M \}.$$

Then, one  has the orthogonality relation: $\forall x \in \Lambda_a$
\be
 \sum_{p \in \widehat \Lambda_a} \exp(\pm ixp) = {\sqrt {2M+1}} \delta_{a, L} (x)\equiv (2M+1) \delta_{x, 0},
\label{Psi_6}
\ee
where $ \delta_{x, 0}$ is the  Kronecker delta.

And, $\forall p \in  \widehat \Lambda_a$,
\be
  \sum_{x \in \Lambda_a} \exp(\pm ixp) = {\sqrt {2M+1}} \delta_{a, L} (p)\equiv (2M+1) \delta_{p, 0}.
\label{Psi_6a}
\ee

Let, $\forall x_1, x_2, x_0 \in \Lambda_a$,
 
 \be
\Psi_{EPR}^{a, L} (x_1, x_2)  =\sum_{p \in \widehat \Lambda_a} \exp(i(x_1-x_2+x_0)p),
\label{Psi_5}
\ee
where the sum $x_1-x_2+x_0$ is modulo $2aM$.

 Using (\ref{Psi_6}),

 \be
\Psi_{EPR}^{a, L} (x_1, x_2)  = \sqrt {2M+1} \delta_{a, L} (x_1-x_2+x_0) 
\label{Psi_7}
\ee
can be written as:

\be
\Psi_{EPR}^{a, L} (x_1, x_2)  =   \sum_{x \in  \Lambda_a} \delta_{a, L} (x-x_2+x_0) \delta_{a, L} (x_1-x).
\label{Psi_7a}
\ee

One can also introduce the finite Fourier transform:
\be
\widehat \Psi (p_1, p_2) =  \frac{1}{ {2M+1}}\sum_{x_1, x_2 \in  \Lambda_a} \exp(-i(x_1p_1+x_2p_2))  \Psi (x_1, x_2)
 \label{Psi_9}
\ee

whose inverse is:
\be
 \Psi (x_1, x_2) = \frac{1}{ {2M+1}} \sum_{p_1, p_2 \in  \widehat \Lambda_a} \exp(i(x_1p_1+x_2p_2))  \widehat \Psi (p_1, p_2).
 \label{Psi_9a}
\ee

The analogues of the operators $P_1$, $P_2$, $Q_1$, $Q_2$ of Subsection \ref{sec3.2} are:
\be
P_j \Psi (x_1, x_2)= \sum_{p_1, p_2 \in \widehat \Lambda_a} \exp(i(x_1p_1+x_2p_2)) p_j \widehat \Psi (p_1, p_2), \quad j=1,2,
 \label{Psi_8}
\ee
and 
\be
Q_j \Psi (x_1, x_2)= x_j  \Psi (x_1, x_2),  \quad j=1,2.
 \label{Psi_10}
\ee

These operators have proper (not generalized) eigenvectors:

\be
P_j \exp(-i(x_1p_1+x_2p_2))= p_j  \exp(-i(x_1p_1+x_2p_2))
 \label{Psi_11}
\ee
and
\be
Q_j \delta_{a, L} (x_1 -x_{0, 1}) \delta_{a, L} (x_2 -x_{0, 2})= x_{0, j}  \delta_{a, L} (x_1 -x_{0, 1})
 \delta_{a, L} (x_2 -x_{0, 2}).
 \label{Psi_12}
\ee

Thus, if one applies the collapse rule for the measurement of the observable $P_1$ to $\Psi_{EPR}^{a, L} (x_1, x_2)$, when the observed value is $p$, the resulting state will be proportional to $  \exp(i(x_1-x_2+x_0)p)$, meaning that the state of the second particle will be  proportional to $\exp(-i(x_2-x_0)p)$.
And,  if one applies the collapse rule for the measurement of the observable $Q_1$ to $\Psi_{EPR}^{a, L} (x_1, x_2)$, when the observed value is $x$, the resulting state will be  proportional to $  \delta_{a, L} (x-x_2+x_0) \delta_{a, L} (x_1-x)$, meaning that the state of the second particle will be  proportional to $\delta_{a, L} (x-x_2+x_0)$.

 \section{Proof of nonlocality using the EPR Variables}\label{sec4}

Given a state like (\ref{Psi_3}, \ref{Psi_4}), we can almost repeat the arguments of Section \ref{sec2} in order to prove nonlocality. First observe that one has an analogue of a Schrödinger theorem.
Consider a generalized state for four particles in one dimension:

\ba
\delta (x_1-x_3+x_0) \delta (x_2-x_4+x_0),
\label{state}
\ea
which is just the product of two copies of the EPR state (up to a $4\pi^2$ factor, see (\ref{Psi_4})), one for the pair of particles $(1,3)$, the other for the pair of particles $(2,4)$. Alternatively, one may regard this as a state of two particles in two dimensions, with coordinates $(x_1, x_2)$ and  $(x_3, x_4)$. In our previous notation, system 1 will consist of particles 1 and 2 and system 2 will consist of particles 3 and 4.\footnote{We need two copies of the EPR state only in order to prove Theorem \ref{3} below.}

One may also replace that state by its regularized version, see (\ref{Psi_7}):

\ba
\delta_{a, L} (x_1-x_3+x_0) \delta_{a, L} (x_2-x_4+x_0).
\label{rstate}
\ea

By Remark 2 in Subsection \ref{sec2.1}, the state (\ref{rstate}) is maximally entangled and so the state (\ref{state})
is also (formally) maximally entangled.\footnote{Formally, since the state itself is not a  vector in a finite dimensional Hilbert space. But, since we are not concerned here with mathematical rigor, we will put aside that issue.}

We  need to introduce standard operators $Q_1$, $Q_2$, $Q_3$, $Q_4$, that act as multiplication on a suitable set of functions in $L^2(\R^4)$:
\ba
Q_j \Psi( x_1, x_2, x_3, x_4)= x_j \Psi( x_1, x_2, x_3, x_4)\;,\quad j=1,2,3,4\;,
\label{Q}
\ea
and operators $P_1$, $P_2$, $P_3$, $P_4$ that act by differentiation on a suitable set of functions in $L^2(\R^4)$:
\ba
P_j \Psi( x_1, x_2, x_3, x_4)= -i\frac{\p}{\p x_j} \Psi( x_1, x_2, x_3, x_4)\;,\quad j=1,2,3,4\;.
\label{P}
\ea
Or, using  the Fourier transform (\ref{FT}) of $\Psi$ (for four variables):
 \ba
&&P_j \Psi( x_1, x_2, x_3, x_4) =\non
\\&&  \frac{1}{(2\pi)^2} \int \exp(i (p_1 x_1+p_2 x_2+p_3 x_3+p_4 x_4))  p_j \widehat \Psi( p_1, p_2, p_3, p_4) dp_1 dp_2, dp_3 dp_4, 
\label{P2}
\ea
$\mbox{for} \quad j=1,2,3,4 $.

Consider the eight operator $Q_1$, $Q_2$, $Q_3$, $Q_4$, $P_1$,  $P_2$, $P_3$,  $P_4$, defined by (\ref{Q}) and (\ref{P}, \ref{P2}). 

 Let $\CB$ be the set of products of analytic functions of one of the operators  $Q_1$, $Q_2$, $P_1$,  $P_2$ defining a self-adjoint operator, and  let $\tilde\CB$ be the set of sums  of products of analytic functions of one of the operators  $Q_3$, $Q_4$, $P_3$,  $P_4$ defining a self-adjoint operator.
 
 Given the maximally entangled state (\ref{state}), 
 for every operator $\tilde O\in \tilde \CB$, there is a corresponding (in the sense of the Principle of Perfect Correlations) operator $O \in \CB$, and vice-versa.
 (For $x_0=0$, $O$ is obtained by changing in $\tilde O$ the index $3$ to $1$ and  the index $4$ to $2$). 
 And, by Schrödinger's theorem,  assuming locality,  there is a  non-contextual value-map $v: { \CB} \rightarrow {\R}$ that satisfies (\ref{res}) and therefore also the property (\ref{res3}).

However this  is contradicted by a theorem of Robert Clifton \cite{RC}, proven in the appendix.

\vspace*{5mm}

\begin{theorem}\label{3}
 {\bf Non-existence of pre-existing values for positions and momenta.}
 
 Consider the set of analytic functions of one of the operators $Q_1$, $Q_2$, $P_1$,  $P_2$. And let $\CB$ be the set of products of such functions defining a self-adjoint operator. Then, there does not exist a map
\ba
v: { \CB} \rightarrow {\R}
\label{res1a}
\ea
such that:
\begin{itemize}
\item[1)]
\ba
v(f( O))=f (v( O)),
\label{res2a}
\ea
for any real valued function $f$ of a real variable.
\item[2)]
$\forall O, O' \in { \CB}$ with  $ [O, O']= OO'-O'O=0$, (\ref{res3}) for $f(x,y)=xy$ holds: 
\be
v(O O')=v(O) v(O').\label{res3a}\\
\ee

\end{itemize}

In particular, there cannot exist a non-contextual value-map.
\end{theorem}

\vspace*{5mm}

 So, combining the EPR argument with the previous theorem, we again establish nonlocality, without using Bell's inequalities.

The logic is the same as in Section \ref{sec2}:

\begin{itemize}
\item [1] EPR show that the perfect correlations plus locality imply that the values of some physical quantities (the values $v(O)$ of the operators $O\in \CA$ in Section \ref{sec2.3} or  the operators $O\in \CB$ here), must exist independently of whether one measures them or
not, and that defines a  non-contextual value-map.

\item [2]  Theorems \ref{2} or \ref{3}  show that merely assuming the existence of such a map
leads to a contradiction.
\end{itemize}

{\it Therefore the locality assumption is false!}

\section{What happens in Bohmian mechanics?}\label{sec5}

In Bohmian mechanics, or pilot-wave theory, the complete state of a closed physical system composed of $N$ particles is a pair 
$(|$quantum state$>$, $\bf X)$, where $|$quantum state$>$ is the usual quantum state (given by the tensor product of wave functions with some possible internal states), and 
${\bf X}= (X_1,\ldots, X_N)$ is the configuration representing the  positions of the particles (that exist, independently of whether one   ``looks" at them or one  measures them; each $X_i\in \R^3$).\footnote{ For elementary introductions to this theory, see \cite{Bri2, Tu} and for more advanced ones, see \cite{B, Bo1, BH, Bri1, DGZ, DT,  DGZ1, Go, No}. There are also pedagogical videos made by students in Munich, available at: https://cast.itunes.uni-muenchen.de/vod/playlists/URqb5J7RBr.html.}

These positions are the  ``hidden variables" of the theory, in the sense that they are not included in the purely quantum description $ |$quantum state$>$, but they are not at all hidden: it is only the particles' positions that one detects directly, in any experiment (think, for example, of the impacts on the screen in the two-slit experiment). So the expression ``hidden variables" is really a misnomer, at least in the context of Bohmian mechanics.

Both objects, the quantum state and the particles' positions, evolve according to deterministic laws, the quantum state guiding the motion of the particles. Indeed,
the time evolution of the complete physical state is composed of two laws (we consider, for simplicity, spinless particles):

\par\noindent
1.  The wave function evolves according to the usual Schr\"odinger's equation.

\bigskip 

\par\noindent
2.
The particle positions ${\bf X}={\bf X}
(t)$  evolve in time 
 according to a guiding equation determined by the quantum state: their velocity is a function of the wave function. If one writes\footnote{We use lower case letters for the generic arguments of the wave function and upper case ones for the actual positions of the particles.}:
$$\Psi (x_1, \dots, x_N)=R (x_1, \dots, x_N)e^{iS (x_1, \dots, x_N)},
$$ then:

 \be
 \frac{ d X_k (t)}{dt}=   \displaystyle  \nabla_k S (X_1(t),\ldots,X_N(t)),
 \label{P0}
 \ee
where $ \nabla_k$ is the gradient with respect the coordinates of the  $k$-th particle.

In order to understand why Bohmian mechanics reproduces the usual quantum predictions, one must use a fundamental consequence of that dynamics, 
{\it equivariance}:  If the probability density $\rho_{t_0}({\bf x})$ for the initial configuration $ {\bf X}_{t_0}$
 is given by $\rho_{t_0}({\bf x}) = |\Psi ({\bf x}, t_0)|^2$, then the probability density for the configuration ${\bf X}_t$ at any time t is given by 
\ba
 \rho_t ({\bf x})= |\Psi ({\bf x}, t)|^2,
  \label{QE}
\ea
where $\Psi  ({\bf x}, t)$ is a solution to  Schr\"odinger's equation.
This follows easily from equation (\ref{P0}).

Because of equivariance, the quantum predictions for the results of   measurements of any quantum observable  are
obtained  if one assumes that the initial density satisfies $\rho_{t_0}  ({\bf x}) = |\Psi  ({\bf x}, t_0)|^2$. The assertion
that configurational probabilities at any time $t_0$ are given by this ``Born rule" is called
the {\it quantum equilibrium hypothesis}. The justification of the quantum equilibrium hypothesis -- and, indeed, a clear  statement of what it actually means -- is a long story, too long to be discussed here (see \cite{DGZ}).

In Bohmian mechanics, particles have a velocity at all times and therefore they have what we would be inclined to call a momentum (mass $\times$ velocity). So one might ask, what sort of probability does Bohmian mechanics supply for the latter: will it agree with the quantum mechanical probability for momentum? The answer, as we will see in the next subsection,  is no!

One may also ask: isn't having both a position and a velocity at the same time contradicted by Heisenberg's inequalities? 
Moreover, since Bohmian mechanics is deterministic,  the result of any quantum experiment must be pre-determined by the initial conditions of the 
system being measured and of the measuring device. But why doesn't that provide a non-contextual value-map whose existence is precluded by Theorem \ref{3}? We will discuss these issues in the following subsections and this will also  provide an example of how nonlocality manifests itself in Bohmian mechanics.

\subsection{The measurement of momentum in Bohmian mechanics}\label{sec5.4}

To understand what is going on, we should analyze ``momentum measurements," i.e., what are called momentum measurements in standard quantum mechanics.
Consider a simple example, namely a particle in one space dimension with initial wave function $\Psi (x, 0)=\pi^{-1/4}\exp(- x^2/2)$. Since this function is real, its phase  $S=0$  and the particle is  at rest (by equation (\ref{P0}): $ \frac{d X(t)}{dt} =  \frac{\partial S(X(t),t)}{\partial x}$).
Nevertheless, the measurement of momentum  $p$ must have, according to the usual quantum predictions,  a probability distribution whose density is given by the square of the  Fourier transform of $\Psi (x, 0)$, i.e. by
 $|\hat \Psi (p)|^2=\pi^{-1/2}\exp(- p^2)$.
 Isn't there a contradiction here?  Isn't there a clear disagreement with the quantum predictions?
 
In order to answer this question, one must focus on the quantum mechanical {\it measurement}
 of momentum. One way to do this is to let the particle move freely and to detect its asymptotic position  $X(t)$ as
   $t\to \infty$. Then, one sets  $p= \displaystyle \lim_{t\to \infty} \frac{X(t)}{t}$
 (putting the mass $m=1$).
 
Consider the free evolution of the initial wave function at $t_0=0$, $ \Psi (x, 0)=\pi^{-1/4}\exp(- x^2/2)$. The solution of  Schr\"odinger's equation with that initial condition is:

\begin{equation}
\Psi (x,t) = \frac{1}{(1+it)^{1/2}} \frac{1}{\pi^{1/4}}\exp\left[- \frac{x^2}{2 (1+it)}\right],
\label{Ga1}
\end{equation}
and thus
\begin{equation}
| \Psi (x,t)|^2=\frac{1}{\sqrt{\pi \big[1+t^2\big]} }\exp\left[- \frac{x^2}{ 1+t^2}\right].
\label{Ga}
\end{equation}

If one writes $\Psi (x,t)= R (x,t) \exp \big[iS (x,t)\big]$, one gets  (up to a $t$-dependent constant): 
\ba
S(x,t)= \frac{t x^2}{2 (1+t^2)}, 
\ea
and the guiding equation (\ref{P0}) becomes:
\begin{equation}
 \frac{d}{dt} X(t)=  \frac{t X(t)}{1+t^2},
\label{Ga2}
\end{equation}
whose solution is:
\begin{equation}
X(t)=X(0) \sqrt{1+t^2}.
\label{Ga3}
\end{equation}
This gives the explicit dependence of the position of the  particle as a function of time.
If the particle is initially at   $X(0)=0$, it does not move; otherwise, it moves asymptotically, when  $t\to \infty$, as $X(t)\sim X(0)t$.
Thus, $p = \lim_{t\to \infty} X(t)/t = X(0)$.

Now, assume that we start with the quantum equilibrium distribution: 
$$
\rho_0 (x)= |\Psi (x, 0)|^2=\pi^{-1/2}\exp(- x^2).
$$
This is the distribution of $X(0)$.
Thus, the distribution of  $p = \lim_{t\to \infty} X(t)/t = X(0)$ will be
$\pi^{-1/2} \exp (-p^2) = |\hat \Psi(p,0)|^2$. This is the quantum prediction! But the detection procedure (measurement of $ X(t)$ for large $t$) does {\it not }  measure the initial velocity (which is zero with probability $1$).

{\bf Remarks}

\begin{itemize}
\item[1.] 
Although the particles do have, at all times, 
 {\it a position and a velocity},  there is no contradiction between
Bohmian mechanics and the quantum predictions and, in particular, with Heisenberg's uncertainty principle. The latter is simply a relation between variances of results of measurements. It implies nothing whatsoever about what exists or does not exist outside of measurements, since those relations  are simply mathematical consequences of the quantum formalism which, strictly speaking, dictates only what takes place during a measurement.

\item[2.] 
Bohmian mechanics shows that what are called measurements of quantum observables other than positions are typically merely {\it interactions} between a microscopic physical system and a macroscopic measuring device whose statistical results coincide with the quantum predictions. 
 
To use a fashionable expression, one might
say for both Bohmian mechanics and standard quantum mechanics,  values of most observables are {\it emergent}. But it is only in Bohmian mechanics that one can understand how that emergence comes about.

 \end{itemize}

\subsection{The contextuality of the momentum  measurements in Bohmian mechanics}\label{sec5.5}

The reader might nevertheless worry that there {\it is} in fact an intrinsic  property of the particle that is revealed in a momentum  measurement, for example  its original position, since, as we  showed in the previous subsection, $p = \lim_{t\to \infty} X(t)/t = X(0)$ in the simple case considered there. Of course, if one were to measure the position one would also find an  intrinsic  property of the particle (namely its position!).\footnote{The fact that  the measurements of both the momentum and the position reveal the same intrinsic  property may sound strange but that is just a peculiarity of  the example considered here.} But doesn't that contradict our Theorem \ref{3} (our example could of course be formulated in two dimensions by taking a product of wave functions of the form (\ref{Ga1}))? After all, the latter theorem asserts that  {\it there does not exist } a value-map that  assigns to a quantum system  pre-existing values that are revealed by quantum measurements and here we seem to have just defined such a map. 

However, as we shall explicitly show, the map provided by Bohmian mechanics would be contextual (see  the Appendix  for the concrete operators that we use in the proof of  Theorem \ref{3}). In particular the value
 $v(O)$  will depend on which other operators $O', O'', \dots$,  one measures together with $O$. Hence relations like (\ref{res3a}) that are needed to prove Theorem \ref{3} will not be valid: for example, if one writes $v(O O') = v(O) v(O')$ and $v(O O'') = v(O) v(O'')$, the value $v(O)$ will in general be different in the two relations. 

We will now show in particular that the measurement of momentum is contextual, using a modified version of the example given by (\ref{Ga1}).
 
Take that quantum state (\ref{Ga1}) and  write $\Psi_0(x$) for $\Psi(x,0)$. Consider the  corresponding Gaussian wave functions:

\begin{equation}
\Psi_{+k}(x) = \Psi_0(x) e^{ikx}
\label{Ga4}
\end{equation}
and
\begin{equation}
\Psi_{-k}(x) = \Psi_0(x) e^{-ikx}
\label{Ga5}
\end{equation}
where $k>0$. We will assume below that $k$ is large.

Consider first the initial wave function $\Psi_{+k}(x) = \Psi_0(x) e^{ikx}$. This is a right-moving Gaussian wave packet moving with speed $k$. Thus at time $t$ it will be centered at $kt$. Explicitly, the solution of Schrödinger's
equation is:

\ba
\Psi_{+k}(x, t) =  \frac{1}{(1+it)^{1/2}} \frac{1}{\pi^{1/4}} \exp\left(ikx-\frac{ik^2t}{2}- \frac{(x-kt)^2 }{2 (1+it)}\right),
\label{Ga5b}
\ea 
which can also been seen immediately from (\ref{Ga1}) using Galilean invariance. For this wave packet we  have that 
$p = \lim_{t \to \infty}  \frac{X(t)}{t}  \approx k$ for $k >> 1$.

 Now form an $N=2$ entangled state $\Psi$ from the wave functions (\ref{Ga4}, \ref{Ga5})\footnote{This state ressembles a maximally entangled one, but it does not fit the definition of maximally entangled, since  the Hilbert space here in infinite dimensional.}:

\begin{equation}
\Psi(x,y) = A[ \Psi_{+k}(x)\Psi_{+k}(y) + \Psi_{-k}(x)\Psi_{-k}(y)],
\label{Ga6}
\end{equation}
with $A$ the normalization constant.
Let $O = P_x$. Consider two different experiments that measure  $O$:

$\mbox{Experiment}_1(O) $: measure $O$ alone 
by the procedure 
described in Subsection \ref{sec5.4}, with result corresponding to the solution to the guiding equation (\ref{P0}) associated with the solution of Schr\"odinger's equation.

$\mbox{Experiment}_2(O) $ : first measure at time $0$ the position $Q_y$ of the second particle, then measure $O$ by the above procedure.

For   $\mbox{Experiment}_1(O)$, we claim that the result is

\be
v(\mbox{Experiment}_1(O))  \approx  \sgn (X(0) + Y(0)) k
\label{Ga7c}
\ee 
for $k$ large.

To prove (\ref{Ga7c}),
introduce the variables:
\ba
\nonumber
w &=&  \frac{x+y}{\sqrt 2}, \\
z &=&  \frac{x-y}{\sqrt 2}.
\label{Ga8}
\ea
In terms of these variables,  we can rewrite (\ref{Ga6}) as
\begin{equation}
\Psi(w, z) =A (\Psi_{+k'}(w) + \Psi_{-k'}(w)) \Psi_0(z).
\label{Ga9}
\end{equation}
with $k' = \sqrt2 k$. 

So the solution of Schr\"odinger's equation  factorizes into a function $\Psi (w, t)$ of $(w, t)$ and 
a function $\tilde \Psi (z, t)$ of $(z, t)$. We have that 
$\tilde \Psi (z, t)$ is given by (\ref{Ga1}) with $x$ replaced by $z$, while for $\Psi (w, t)$ we get a
sum of two wave functions like (\ref{Ga5b}), one with $k$ replaced by $k'$, the other with $k$ replaced by $-k'$:
\begin{equation}
\Psi(w, t) =A(\Psi_{+k'}(w, t) + \Psi_{-k'}(w,t))
\label{Ga9a}
\end{equation}
with $\Psi_{\pm k'}(w, t)$ of the form (\ref{Ga5b}).

For large $t$, $|\Psi (w, t)|^2$ is a sum of two more or less non-overlapping  terms, one corresponding to the part of the wave function with $k'$ (whose support is around $k't$), the other one corresponding to the part of the wave function with $-k'$ (whose support is around $-k't$): 

\begin{equation}
|\Psi (w, t)|^2 \approx A^2 (|\Psi_{+k'}(w, t)|^2 + |\Psi_{-k'}(w,t)|^2).
\label{Ga9b}
\end{equation}

Since the solution of Schr\"odinger's equation  factorizes into a function of $(w, t)$ and one of $(z, t)$,  the guiding equations (\ref{P0}) for $W(t)$ and $Z(t)$ are decoupled. 
For $Z(t)$ we obtain a solution like (\ref{Ga3}) ($Z(t) \approx Z(0)t$ as $t \to \infty$).

To analyze $W (t)$, note that one property of the dynamics (\ref{P0}) is that, in one
dimension, trajectories cannot cross.\footnote{That is because there is a unique solution of the first order equation (\ref{P0}) if the position is fixed at a given time.} Since there is a symmetry between the two parts of the wave function
(\ref{Ga9a}) (upon reflection, $\Psi_{+k'}$ becomes $\Psi_{-k'}$ ), if the initial condition $W (0) > 0$, the particle must stay on the right, while if
$W (0) < 0$, the particle must stay on the left. Moreover, by equivariance, the particle evolves so
as to be in the support of $|\Psi (w, t)|^2$, which, by  (\ref{Ga9b}), consists of two non-overlapping
terms supported around $\pm k't$  for large t.  So, for large $k$ and large times, we get that $W (t) \approx \sgn W (0)k' t = \sgn W (0) \sqrt2 kt$.

Rewriting what we've found in terms of the $X(t)$ and $Y(t)$ variables, we get that $X (t)= \frac{W(t)+Z(t)}{\sqrt 2} \approx \frac{1}{\sqrt 2}( \sgn W (0) \sqrt2 k t + Z(0)t)$ and thus, $v(\mbox{Experiment}_1(O))= \lim_{t \to \infty} \frac{X(t)}{t}  \approx \sgn (X(0) + Y(0)) k$, for $k$ large, which is  (\ref{Ga7c}).

For $\mbox{Experiment}_2(O) $,  if $Y$ is the result of the measurement of $Q_y$,
the wave function (\ref{Ga6}) collapses, yielding for the wave function of the $x$ system\footnote{In Bohmian mechanics,  in fact, there  is an actual collapse of the (conditional) wave function of a system upon measurement; see \cite[Sect. 6.1]{BH}, \cite{Be6}.}:
 \be 
\Psi(x)=A(Y)  (c_+(Y) \Psi_{+k}(x) +  c_-(Y)\Psi_{-k}(x)).
 \label{Ga10}
\ee
 with $c_\pm (Y) =\Psi_{\pm k}(Y)$ and $A(Y)$ the normalization coefficient.
  
  The solution of  Schr\"odinger's equation with this initial condition is again a sum  of  two wave functions like (\ref{Ga5b}), one with $+k$, the other with $-k$, multiplied by coefficients $c_\pm (Y)$:
\be 
\Psi(x, t)= A(Y)(c_+(Y) \Psi_{+k}(x, t) +  c_-(Y)\Psi_{-k}(x, t)),
 \label{Ga11}
\ee
where $\Psi_{\pm k}(x, t)$ of the form (\ref{Ga5b}).

We can now more or less reason as we just did for the $\Psi (w, t)$ given by (\ref{Ga9a}), except
that because of the coefficients $c_\pm (Y)$ there is no symmetry here between the two parts of the wave function---unless the complex exponentials in $c_\pm (Y)$ are real (i.e. $e^{ikY}=\pm 1$). Nonetheless, the effect of the coefficients in (\ref{Ga11}) is merely to replace the $\cos kx$, which would arise there if $c_\pm (Y)>0$ (i.e. $e^{ikY}= 1$), by its translate $\cos(kx+kY)$. Thus the $|\Psi|^2$  probability of the interval $[X_m, \infty)$ will be 1/2 for some $X_m$ with $|X_m| < \frac{\pi}{2k}$.\footnote{In fact, $X_m$  must lie between $0$ and the nearest maximum of $\cos^2(kx + kY)$.} Thus, by no-crossing and equivariance, we get that for large times $X(t) \approx \sgn(X(0)) kt $ for $k  >> 1$, and thus

\be
v(\mbox{Experiment}_2(O))=  \lim_{t \to \infty} \frac{X(t)}{t} \approx \sgn (X (0)) k. 
\label{Ga7b}
\ee

Comparing (\ref{Ga7c}) and  (\ref{Ga7b}), we see that   the measurement of momentum is contextual, since it may depend on whether or not one measures  another operator $Q_y$ together with
$O = P_x$.

\subsection{An example of nonlocality in  Bohmian mechanics  }\label{sec5.6}

It would go far beyond the scope of this paper to really explain how nonlocality appears in Bohmian mechanics in general, but we saw 
 an example of nonlocality in Bohmian mechanics in the previous subsection:  the particles with coordinates $x$ and $y$ having the entangled quantum state (\ref{Ga6}), can be (in principle) as far apart as one wants and the result of the measurement of $ O=P_x$ will depend on whether or not one measures $Q_y$ before measuring $P_x$, and, since the time interval between these two measurements can be arbitrarily small, we have indeed here an example  of an instantaneous action at a distance. Here we should regard the measurement of $P_x$ as taking a (large but) finite time, and $x$ and $y$ as referring to different (distant) origins.

The fact that Bohmian mechanics is nonlocal is obviously a merit rather than a defect, since we know that any  theory accounting for the quantum phenomena must be nonlocal, as shown in Sections \ref{sec2}-\ref{sec4} (and many other places). 

\section{Summary and conclusions}\label{sec6}

Both EPR and Schrödinger argued that the quantum mechanical description of a system by its wave function is incomplete in the sense that other variables must be introduced in order to obtain a complete description. Their argument was very simple: if I can determine the result of a measurement carried at one place by doing another measurement far away from that place, then that result must pre-exist  its measurement. The wave function alone does not tell us what that result is. Therefore, the quantum mechanical description of a system by its wave function is incomplete.

However, there was a crucial assumption in the reasoning of EPR and Schrödinger, which was too obvious 
for them to question it: that doing a measurement at one place cannot possibly affect instantaneously the physical situation far away, or what is now called the assumption of locality.

The history of the EPR--Schrödinger argument is complicated, because although their conclusion about incompleteness of quantum mechanics was right, their assumption of locality was not. The completion of quantum mechanics was found by de Broglie in 1927 and
developed  by Bohm in 1952.
 Bohm showed that one may consistently assume that particles have trajectories and explained on that basis how to understand measurements as consequences of the theory and not, as they are in ordinary quantum mechanics,
 as a {\it deus ex machina}  \cite{Bo1}. 
 
 The falsity of the locality assumption was shown by John Bell in 1964  \cite{Be2} and by subsequent experiments. Bell first recalled that, if one assumes locality, then, as the EPR argument correctly showed, there must exist other variables than the quantum state to characterize a physical system. But then Bell showed that the distribution of those variables must satisfy some contraints that are violated by quantum predictions, predictions that were later verified experimentally (see \cite{GNTZ} for a survey).  

Here and in \cite{1} we give a simpler argument, but using the maximally entangled states introduced by Schrödinger: for those states, one can, for each observable associated to  one system, construct another observable associated to the second system, possibly far away from the first one, such that the results of the measurement of both observables are perfectly correlated. Then, assuming locality, those results must pre-exist  their measurement. But  assuming that, in general,  observables have values before their measurement leads to a contradiction. Hence, the assumption of locality is false.

The difference between this paper and  \cite{1}  is that here we use the position and momentum variables used by EPR, while in  \cite{1} we used spin variables, such as those in terms of which the EPR argument was reformulated by Bohm \cite{Bo}.

Next one might ask how Bohmian mechanics deals with this impossibility of the pre-existence of measurement results prior to measurements, since it is a deterministic theory, and in such a theory everything is pre-determined by the initial condition. In \cite{1} we reviewed that the measurements of spin variables are contextual, in fact should not properly be called measurements at all. Here we illustrate the contextuality of momentum. In both cases, the contextuality is linked to nonlocality, as it must be, since as explained here and in \cite{1}, if locality were true, then measurements must (sometimes) be non-contextual. Bohmian mechanics is an extremely natural version of quantum mechanics, involving the obvious ontology evolving the obvious way. A proper appreciation of the role of contextuality in Bohmian mechanics can help dispel the widespread uneasy feeling that somehow there must be something amiss in that theory.

\begin{appendices}
 
\section{Appendix: Proof of Clifton's Theorem \ref{3}}\label{app2}

\noindent The proof we give here  is taken from a paper by Wayne Myrvold \cite{My}, which is a simplified version of the result of Robert Clifton \cite{RC} and is similar to proofs of David Mermin \cite{Me4}, and to Asher Peres \cite{Per, Per1} in the case of spins. We note that the same proof would apply to the regularized EPR state of Subsection \ref{sec3.6}.

{\bf Proof of Theorem \ref{3}}
 
We will need the operators $U_j (b)= \exp(-i b Q_j)$, $V_j (c)= \exp(-i c P_j)$, $j=1, 2$, with $Q_j$,
$P_j$ defined by formulas (\ref{Q}), (\ref{P}), but acting in $L^2(\R^2)$ instead of $L^2(\R^4)$, and $b, c \in \R$. They act as 
\ba
U_j (b)\Psi( x_1, x_2)=\exp(-i b x_j) \Psi( x_1, x_2)\;,\quad j=1,2\;,
\label{U}
\ea
which follows trivially from (\ref{Q}), and 
\ba
V_1 (c) \Psi( x_1, x_2)=  \Psi( x_1-c, x_2)\;,
\label{V}
\ea
and similarly for $V_2 (c)$. Equation (\ref{V}) follows from (\ref{P}) by expanding both sides in a Taylor series, for functions $\Psi$ such that the series converges, and by extending the unitary operator $V_2 (b)$ to more general functions $\Psi$ (see, e.g., \cite[Chap.~8]{RS1} for an explanation of that extension).

We choose now the following functions of the operators $Q_i$, $P_i\,$:
\ba
A_1 = \cos (a Q_1)\;,\quad A_2 = \cos (a Q_2) \;,\quad B_1 = \cos \frac{\pi P_1}{a}\;,\quad B_2 = \cos \frac{\pi P_2}{a}\;,
\ea
where $a$ is an arbitrary constant, and the functions are  defined by (\ref{U}), (\ref{V}), and the Euler relations:
\begin{equation}\label{A.43b}
\begin{array}{l}
\displaystyle\cos (a Q_j) =\frac{ \exp(i a Q_j)+ \exp(-i a Q_j)}{2}\;,\\
\displaystyle\cos  \frac{\pi P_j}{ a} =\frac{ \exp( i\pi P_j/ a)+ \exp(- i\pi P_j/ a)}{2}\;, 
\end{array}
\end{equation}
for $j=1, 2$. Note that $A_1,A_2, B_1,B_2$ are self-adjoint. By applying  (\ref{res3a}) several times to pairs of commuting operators made of products of such operators, we will derive  a contradiction.  

We have the relations
\begin{equation}
[A_1,A_2]= [B_1,B_2]=[A_1,B_2]=[A_2,B_1]=0\;,
\label{A.42}
\end{equation}
since the relevant operators act on different variables.

We can also prove:
\begin{equation}
A_1B_1 = -B_1A_1\;,\qquad A_2B_2 = -B_2A_2\;.
\label{A.43}
\end{equation}
To show (\ref{A.43}), note that, from (\ref{U}) and (\ref{V}), one gets
\ba
U_j (b) V_j (c)= \exp(-ibc) V_j (c)U_j (b)\;,
\label{UV}
\ea
for $j=1,2$, which, for $bc=\pm \pi$, means
\ba
U_j (b) V_j (c)= - V_j (c)U_j (b)\;.
\label{UV1}
\ea
Now use (\ref{A.43b}) to expand the product   $\cos (a Q_j) \cos (\pi P_j/ a)$, for $j=1,2$, into a sum of four terms;
each term will have the form of the left-hand side of (\ref{UV}) with $b=\pm a$, $c= \pm \pi / a$, whence $bc=\pm \pi$. Then applying (\ref{UV1}) to each term proves (\ref{A.43}).

The relations (\ref{A.42}) and (\ref{A.43}) imply that
\be
[A_1A_2, B_2 B_1]= 0
\label{A.43a}
\ee
since two anticommutations (\ref{A.43}) suffice to move the $B$'s to the left of the $A$'s. Similarly we have that
\be
[A_1B_2, A_2 B_1]= 0.
\label{A.43a'}
\ee
We also have, using  (\ref{A.43}) once, that 
\be
A_1A_2 B_2 B_1=-A_1 B_2 A_2 B_1.
\label{A.43c}
\ee
Thus, with $C= (A_1 A_2)(B_2 B_1)$ and $D= (A_1 B_2)(A_2 B_1)$, we have that
\be
C=-D.
\label{A.43d}
\ee

Now suppose there is a value map $v$ as described in Theorem \ref{3}. Then, from (\ref{res2a}) with $f(x)=-x$, we have that
\be
v(C)=-v(D).
\label{A.43e}
\ee
But by (\ref{A.42}), (\ref{A.43a}) and (\ref{A.43a'}), we also have, by (\ref{res3a}), that
\ba
v(C)=v(A_1 A_2) v(B_2 B_1)= v(A_1)v(A_2)v(B_2)v(B_1)
\label{A.44}
\ea
and
\ba
v(D)=v(A_1 B_2) v(A_2 B_1)= v(A_1)v(B_2)v(A_2)v(B_1).
\label{A.45}
\ea
Thus $v(C)= v(D)$. This is a contradiction unless $v(C)=0$, i.e. unless at least one of $v(A_i)$, $v(B_i)$, $i=1,2$ vanishes.
But, by (\ref{res2a}),
\ba
v(A_i) =\cos (a v(Q_i))
\nonumber
\ea
and 
\ba
v(B_i)=\cos (\frac{\pi}{a} v(P_i)),\nonumber
\ea
and thus $a$ can be so chosen that $v(A_i)$ and $v(B_i)$ are all nonvanishing.
\hfill $\blacksquare$

\end{appendices}

 \end{document}